# CERC - Circular e$^+$e$^-$ Collider using Energy-Recovery Linac

Vladimir N Litvinenko[1,2], Nikhil Bachhawat[1], Maria Chamizo-Llatas[3],
Francois Meot[2] and Thomas Roser[2]

[1] Department of Physics and Astronomy, Stony Brook University, Stony Brook, NY, USA

[2] Collider-Accelerator Department, Brookhaven National Laboratory, Upton, NY, USA

[3] Nuclear and Particle Physics Directorate, Brookhaven National Laboratory, Upton, NY, USA

We present a Circular Energy Recovery Collider (CERC) as an alternative approach for a high-energy high-luminosity electron-positron collider to current designs for high-energy electron-positron colliders either based on two storage rings with 100 km circumference or two large linear accelerators. Using Energy Recovery Linacs (ERL) located in the same-size 100 km tunnel would allow a large reduction of the beam energy losses, and therefore a reduction of the power consumption, while providing higher luminosity. It also opens a path for extending the center-of-mass (CM) energy to 500 GeV, which would enable double-Higgs production, and even to 600 GeV for $t\bar{t}H$ production and measurements of the top Yukawa coupling. Furthermore, this approach would allow recycling of not only the energy but also of the particles. This feature opens the possibility for colliding fully polarized electron and positron beams.

## I. The CERC concept

This paper is an update to the original proposal [1,2] to recycle both energy and particles in a future polarized electron-position collider in order to expand the CM energy reach up to 600 GeV, while increasing the attainable luminosity, and concentrates on developments since the last publications. The Circular Energy Recovery Collider (CERC) performance is shown in Fig. 1 for different values of the RF power consumption. The luminosity of CERC scales proportionally to the consumed RF power.

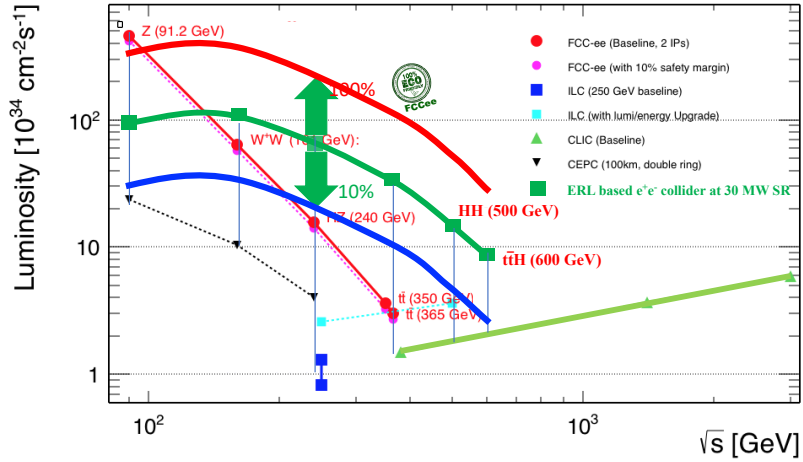

Fig. 1. Luminosities for various options for high-energy e$^+$e$^-$ colliders. We took plot from reference [3] and added three CERC luminosity curves, blue, green and red, for correspong elevels of synchroton radion power: 10 MW, 30 MW and 100 MW.

The CERC design is based on energy recovery linacs (ERL) and two damping rings that are also used for particle recycling. It would consume about one third of power while providing significantly higher luminosity when compared to the SR e$^+$e$^-$ collider, with the only exception at the Z-pole. It will also extend the CM energy reach to 600 GeV, required for the double-Higgs production in the ZHH channel as well as $t\bar{t}H$ production. Even with 30% energy consumption the integrated luminosity per year, at a center-of-mass energy of 500 GeV, would be about 1.5 ab$^{-1}$.

In the CERC design the electron and positron beams are accelerated to the collision energy in 4-path ERL. Most of the energy of the used beams is recovered by delaying them by half of an RF oscillation period and decelerating them. The electron and positron beams are then reinjected into a damping ring, where they are cooled to low emittance prior to repeating the trip. The small amount of beam lost during the process, for example due to scattering from residual gas or burn-off in the collisions, can easily be replaced by adding particles from a linear injector into the electron and positron damping rings. Table 1 lists the main parameters for CERC operating at various beam energies.

Table 1. Main parameters of ERL-based $e^+e^-$ collider with synchrotron radiation power of 30 MW.

| CERC | Z | W | H(HZ) | ttbar | HH | Httbar |
|---|---|---|---|---|---|---|
| Circumference, km | 100 | 100 | 100 | 100 | 100 | 100 |
| **Beam energy, GeV** | **45.6** | **80** | **120** | **182.5** | **250** | **300** |
| Hor. norm ε, μm rad | 3.9 | 3.9 | 6.0 | 7.8 | 7.8 | 7.8 |
| Vert. norm ε, nm rad | 7.8 | 7.8 | 7.8 | 7.8 | 7.8 | 7.8 |
| Bend magnet filling factor | 0.9 | 0.9 | 0.9 | 0.9 | 0.9 | 0.9 |
| $\beta_h$, m | 0.5 | 0.6 | 1.75 | 2 | 2.5 | 3 |
| $\beta_v$, mm (matched) | 0.2 | 0.3 | 0.3 | 0.5 | 0.75 | 1 |
| Bunch length, mm | 2 | 3 | 3 | 5 | 7.5 | 10 |
| Charge per bunch, nC | 13 | 13 | 25 | 23 | 19 | 19 |
| Ne per bunch, $10^{11}$ | 0.78 | 0.78 | 1.6 | 1.4 | 1.2 | 1.2 |
| Bunch frequency, kHz | 297 | 270 | 99 | 40 | 16 | 9 |
| Beam current, mA | 3.71 | 3.37 | 2.47 | 0.90 | 0.31 | 0.16 |
| Luminosity, $10^{35}$ cm$^{-2}$sec$^{-1}$ | 6.7 | 8.7 | 7.8 | 2.8 | 1.3 | 0.9 |
| Energy loss, GeV | 4.0 | 4.4 | 6 | 17 | 48 | 109 |
| Rad. power, MW/beam | 15.0 | 14.9 | 14.9 | 15.0 | 16.8 | 16.9 |
| ERL linacs, GV | 10.9 | 19.6 | 29.8 | 46.5 | 67.4 | 89 |
| Disruption, $D_h$ | 2.2 | 1.9 | 0.8 | 0.5 | 0.3 | 0.3 |
| Disruption, $D_v$ | 503 | 584 | 544 | 505 | 459 | 492 |

The two main effects that affect the luminosity – or quality of collisions - in linear colliders, beamstrahlung and beam disruption, have been studied for the CERC concept. It is expected that large disruption parameters would result both in pinching of the beam sizes as well as in transverse emittance growth. We conducted preliminary studies of these effects in a strong-strong collision simulations and showed that growth of vertical emittance is limited to ~ 4-fold for selected disruption paratmeters.

The main challenge of extending CERC operation to energies above 182.5 GeV is the low energy tail in recirculating electron genearated by beamstrahlung, e.g. high energy photons generated during beam's collisions. Critical energy of the beamstrahlung photons can reah 1 GeV for CERC operation with 300 GeV beams. We selected CERC's energy recovery and damping ring system with an energy acceptance that exceeds the energy of beamstrahlung photons by 10-fold. Such a choice guarantees that beam losses will be less than 1 ppm. Electron and positron bunches are decompressed – up to 15-fold for 300 GeV CERC operaions – in the low energy pass of the ERL prior to injection into the damping rings. The operating energy of the damping rings depends on the CERC top beam energy: it is 2 GeV for CERC beam energies up to 120 GeV, 3 GeV for 182.5 GeV, 4.5 GeV for 250 GeV and 8 GeV for 300 GeV beams. The combination of bunch decompression with ±5% energy acceptance of the damping rings would ensure nearly perfect recovery of the circulating particles. We plan to do Monte-Carlo simulations to identify the exact number of particles that could be lost.

### II.     Studies of the collider lattice.

With diffusion caused by quantum fluctuations of the synchrotron radiation scaling as the seventh power of the beam energy, preservation of the transverse emittance in the accelerating beams is most challenging for the highest proposed energy of operation of 250-300 GeV. We found that using a FODO lattice with a 16-meters period (e.g., two 8-meter combined function magnets) and a phase advance of 90 degrees can satisfy the requirements specified in Table 1 [2]. Conditions for the lower collision energies can also be satisfied.

The lattice of each path around 100-km circumference is comprised of 6250 FODO cells with combined function (dipole, quadrupole, and sextupole) magnets and zero chromaticity. The cell comprises of two 7.6-meter-long magnets separated by 0.4 mm drifts. At the top energy, the combined magnets have a magnetic field of 0.0551 T, field gradients of ±32.24 T/m, and sextupole components SF=267 T/m$^2$ and SD=-418 T/m$^2$. The magnets have an aperture of ±0.75 cm and pole-tip fields of about ~ 1 kG, which is convenient for magnetic steel. Optics functions and emittance evolution of this lattice are show in Fig.2.

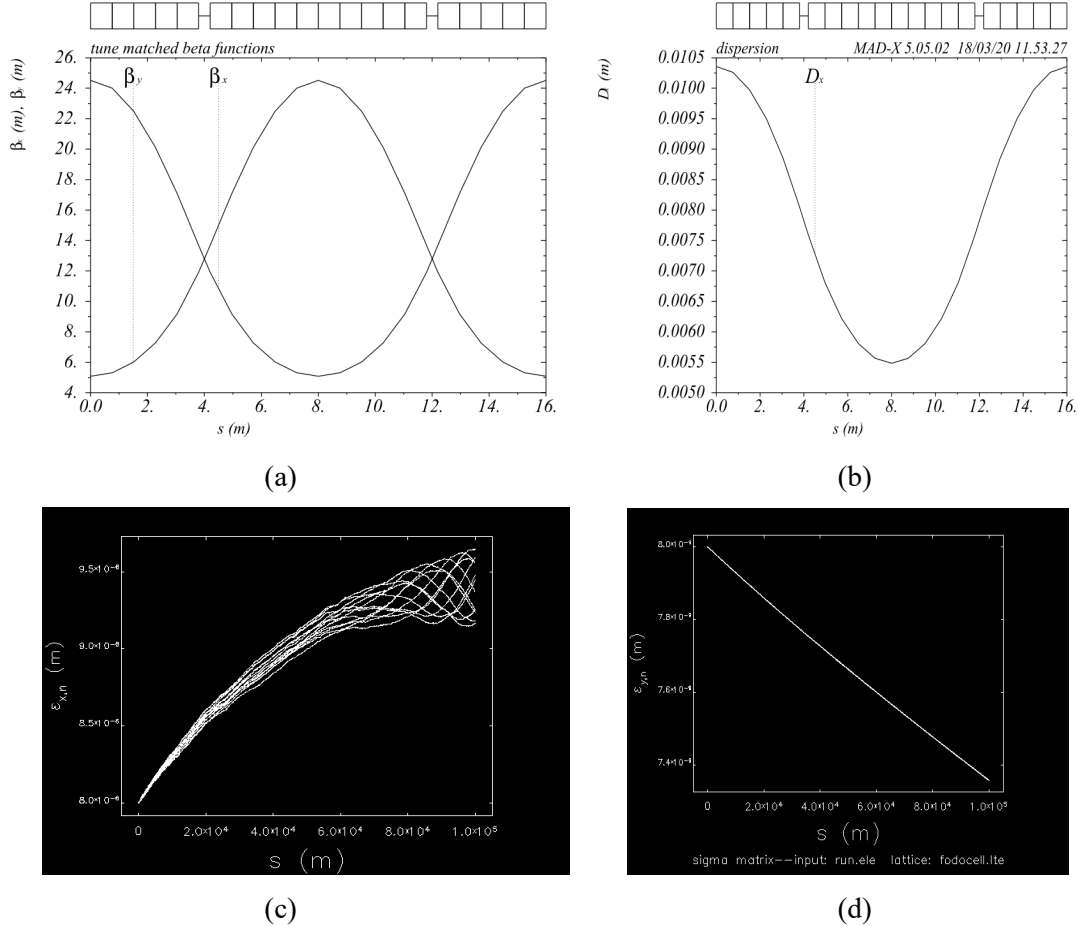

Fig. 2 (a) β- and (b) dispersion- function of the regular ERL lattice. Evolution of normalized horizontal (c) and vertical emittances in the 100 km pass at the top energy.

The electron and positron beams need to undergo compression during the first pass around the tunnel as well as decompression during the last pass prior to reinjection into the damping rings. Long bunches, requiring subsequent compression, have a relatively low peak current in the damping rings, which minimizes Intra-Beam Scattering (IBS). Similarly, the decompression will reduce the energy spread accumulated by the bunches during the acceleration/collision/deceleration cycle, which then can fit into the energy acceptance of the damping rings. Using the low energy passes of the ERL for the compression and decompression will provide for a large value of the longitudinal dispersion $R_{56}$, while maintaining low emittance growth. This process will require additional RF gymnastics, such as chirping beam energy and compensating the energy chirp after the bunch compression/decompression. We are pursuing a detailed design of the entire accelerator, including compressing and de-compressing arcs, SRF linac with splitters and combiners. Details of this studies will be published elsewhere.

One of the advantages of CERC is its capability of colliding polarized beams. Our preliminary studies using the ZGOUBI code confirmed that the proposed lattice can preserve polarization. The simulations include effects of synchrotron radiation, beam emittances and orbit misalignments [4]. The conclusions of these studies is that beam depolarization does not exceed 0.1% per path and that collisions of highly polarized beam in such collider might be feasible.

The impact of different polarization values for electrons and positrons on the production cross section for ZH, ZHH and ttH, has been estimated with Madgraph [5] and it is shown in Table 2. The proper combination of polarization for electrons and positrons will significantly enhance the production cross section or will suppress it.

Table 2: Impact of polarization in the ZH, ZHH, ttH production cross sections.

| Polarization | | Scaling factor | | |
|---|---|---|---|---|
| e $^-$ | e $^+$ | ZH(240GeV) | ZHH(500GeV) | ttH(600GeV) |
| Unpolarized | | 1. | 1. | 1. |
| -70 | 0 | 1.15 | 1.15 | 1.23 |
| -70 | +50 | 1.61 | 1.61 | 1.87 |
| -70 | -50 | 0.69 | 0.69 | 0.73 |
| -70 | +70 | 1.78 | 1.79 | 2.07 |
| -70 | -70 | 0.51 | 0.51 | 0.51 |
| -50 | +50 | 1.47 | 1.47 | 1.69 |
| +50 | -50 | 1.03 | 1.03 | 0.82 |
| +70 | 0 | 0.85 | 0.85 | 0.69 |
| +70 | +50 | 0.60 | 0.60 | 0.56 |
| +70 | -50 | 1.09 | 1.09 | 0.83 |
| +70 | +70 | 0.51 | 0.51 | 0.51 |

### III. Key technical details and assumptions of the concept.

In this section we provide key technical details, to illustrate feasibility of such a collider. First, we base our assumption for the CERC linac on the shunt impedance of the operational 703 MHz 5-cell cavity (so called BNL-3 design) and progress at FNAL in reaching quality factors of $Q_o=10^{11}$ using novel doping techniques and precise demagnetization of cavity prior to cool-down [6-8].

The BNL-3 5-cell SRF cavity unit has a length of 1.58 m with about 1 meter of accelerating structure. We propose to use 16-meter-long cryostats housing 10 five-cell cavities. We assume that cryomodules will be separated by 1 meter, which corresponds to a 58.8% filling factor for the accelerating field. Room temperature HOM couplers, installed at the ends of each of the cavities, will dissipate the majority of the HOM power. The very high frequency components of HOM power will propagate through the large apertures of the 5-cell linacs and will be absorbed by ferrite-type room temperature HOM absorbers installed between cryomodules. Fig. 3 (a) shows the number of 5 cell cavities as function CERC energy.

The BNL-3 cavity has a HOM loss factor of 0.16 V/pC and 0.12 V/pC for electron bunches with RMS bunch length of 30 mm and 50 mm, respectively. It means that for maximum charge per bunch of 25 nC and 30 mm RMS bunch length particles will lose 4 keV per cavity. Each particle passes through each cavity 8 times: 4 times on the way up and 4 times on the way down in energy. The total loss of the particle's energy into HOM modes ranges from 1.1 MeV for the lowest beam energy to 10.6 MeV for the CERC's top beam energy. The total HOM power losses in the absorbers from the 8 passes by both electron and positions beams does not exceed 250 kW and the maximum power per HOM absorber does not exceed 160 W.

At the top energy for FCC-ee of 182.5 GeV, CERC will require two 2-km-long SRF linacs and 20 MW cryo-plant. HOM losses will be at the 100 kW level. Naturally, going above 182.5 GeV will require longer linacs and a more powerful cryo-plant.

For briefness, we do not describe here second harmonic cavities, which do compensate losses for synchrotron radiation, and other harmonic cavities. It is our estimation that these cavities will contribute between 10% and 20% to the cryo-plant power and the HOM losses.

Naturally, 30 MW of RF power will be need for CERC's second harmonic and damping ring RF systems to compensate for the energy lost to synchrotron radiation.

Microphonics represents an additional challenge for modern SRF ERLs by requiring additional RF power to keep cavities in compliance. Typically, a 20-MV 5-cell cavity may require a transmitter with power ~ 20 kW for this purpose. For CERC operating with 182.5 GeV beams it would require additional 46 MW of RF power, which would greatly reduce the attractiveness of this concept.

Fortunately, CERN successfully tested ferro-electric tuners for SRF cavities, which reduce power requirements to about 200 watts [9]. We assume that this technology will be available for CERC and allocate 3 kW of RF power to compensate for microphonics.

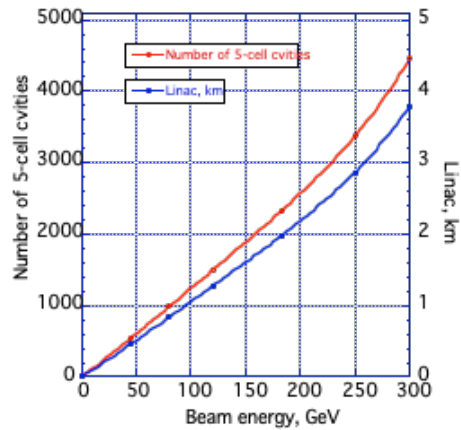

Fig. 3. Number of required 5-cell cavities and length of each SRF linacs as function of the colliding beam energy.

Because in CERC the vertical beam size is measured in microns, we propose to use magnets with a small gap of about 15 mm [2]. Power consumption of the magnet is proportional to its gap and magnetic field squared, which means that 16 beam-lines required for CERC electron and position beams will consume the same amount of power as a single storage ring with typical gap ~ 5 cm. This is why we assumed that the CERC magnetic system will consume about 50% of the power required for the collider storage ring magnets in FCC ee.

### IV.   Summary of CERC power consumption

Finally, Table 3 summarizes our estimation of the CERC power consumption. We are assuming 1.25 kW of cryo-plant power per 1 W loss at 1.8 K in the SRF linac. This includes a 25% overhead related to the cryogenic facility and liquid He transport system. We are also using a ratio of AC power to RF power for the RF amplifiers of 1.66. For the damping rings we would use permanent magnets as is being done now for light sources. The same would be done for the transfer lines to and from the damping rings.

Note that the electric power consumption of CERC is lower by about 100 MW than FCC-ee over its energy range with much higher luminosities at the higher energies. And there is the possibility of extending the center-of-mass energy up to 600 GeV without excessively more power consumption. Electric power consumption could be further reduced with focussed R&D.

Table 3. Estimation of the CERC AC power consumption

| Mode | Beam Energy [GeV] | SR power [MW] | Microphonics [MW] | HOM [MW] | Total RF power [MW] | Magnet [MW] | 1.8K Cryo load [kW] | Cryoplant AC power [MW] | Total AC power [MW] |
|---|---|---|---|---|---|---|---|---|---|
| Z | 45.6 | 30.0 | 1.6 | 0.1 | 31.7 | 2.0 | 5 | 6.25 | 61 |
| W | 80 | 30.0 | 2.9 | 0.2 | 33.1 | 6.2 | 10 | 12.5 | 74 |
| HZ | 120 | 30.0 | 4.5 | 0.3 | 34.8 | 13.9 | 15 | 18.75 | 90 |
| ttbar | 182.5 | 30.0 | 7.0 | 0.2 | 37.2 | 32.0 | 23 | 28.75 | 123 |
| HHZ | 250 | 30.0 | 10.1 | 0.1 | 40.2 | 60.1 | 34 | 42.5 | 169 |
| Httbar | 300 | 30.0 | 13.4 | 0.0 | 43.4 | 86.6 | 45 | 56.25 | 215 |

### V.    Conclusions and acknowledgements.

We did not find any show-stoppers preventing our design to work as a next generation high-energy polarized e+e- collider. We continue the detailed in-depth study to fully authenticate this ERL-based concept.

Authors would like to thank Dr. Frank Zimmerman and FCC-design team at CERN for opportunity to present and discuss this option at their working meeting. Vladimir Litvinenko would like to acknowledge support by NSF grant PHY-1415252 "Center for Science and Education at Stony Brook University".